\documentclass[12pt]{article}
\usepackage{amssymb}

\usepackage{amsmath}
\usepackage{graphicx}

\newcommand{\ah}{\frac{1}{2}}

\begin{document}
\begin{flushright}
\hfill{ USTC-ICTS-08-08}\\
\end{flushright}
\vspace{4mm}

\begin{center}

{\Large \bf Ricci Dark Energy in Brans-Dicke theory}

\vspace{8mm}

{\large Chao-Jun Feng}

\vspace{5mm}

{\em
 Institute of Theoretical Physics, CAS,\\
 Beijing 100080, P.R.China\\
 Interdisciplinary Center of Theoretical Studies, USTC,\\
 Hefei, Anhui 230026, P.R.China
 }\\
\bigskip
fengcj@itp.ac.cn
\end{center}

\vspace{7mm}

\noindent A holographic dark energy from Ricci scalar curvature called Ricci dark energy was proposed recently. In this
model the future event horizon area is replaced by the inverse of the Ricci scalar curvature. We study the evolution of
equation of state of the Ricci dark energy and the transition from decelerated to accelerated expansion of the universe
in the Brans-Dicke theory, which is a natural extension of general relativity. We find that the current acceleration of
our universe is well explained.

\newpage

\section{Introduction}

Why cosmological constant observed today is so much smaller than the Planck scale? This is one of the most important
problems in modern physics. In history, Einstein first introduced the cosmological constant in his famous field
equation to achieve a static universe in 1917. After the discovery of the Hubble's law, the cosmological constant was
no longer needed because the universe is expanding. Nowadays, the accelerating cosmic expansion first inferred from the
observations of distant type Ia supernovae \cite{Riess:1998cb} has strongly confirmed by some other independent
observations, such as the cosmic microwave background radiation (CMBR) \cite{Spergel:2006hy} and Sloan Digital Sky
Survey (SDSS) \cite{:2007wu}, and the cosmological constant comes back as a simplest candidate to explain the
acceleration of the universe in 1990's.

Holographic principle \cite{Bousso:2002ju} regards black holes as the maximally entropic objects of a given region and
postulates that the maximum entropy inside this region behaves non-extensively, growing only as its surface area. Hence
the number of independent degrees of freedom is bounded by the surface area in Planck units, so an effective field
theory with UV cutoff $\Lambda$ in a box of size $L$ is not self consistent, if it does not satisfy the Bekenstein
entropy bound \cite{Bekenstein:1973ur} $ (L\Lambda)^3\leq S_{BH}=\pi L^2M_{pl}^2 $, where $M_{pl}^{-2}\equiv G $ is the
Planck mass and $S_{BH}$ is the entropy of a black hole of radius $L$ which acts as an IR cutoff. Cohen et.al.
\cite{Cohen:1998zx} suggested that the total energy in a region of size $L$ should not exceed the mass of a black hole
of the same size, namely $ L^3\Lambda^4\leq LM_p^2 $. Therefore the maximum entropy is $S^{3/4}_{BH}$. Under this
assumption, Li \cite{Li:2004rb} proposed the holographic dark energy as follows
\begin{equation}\label{li}
    \rho_\Lambda = 3c^2M_p^2 L^{-2}
\end{equation}
where $c^2$ is a dimensionless constant. Since the holographic dark energy with Hubble horizon as its IR cutoff does
not give an accelerating universe \cite{Hsu:2004ri}, Li suggested to use the future event horizon instead of Hubble
horizon and particle horizon, then this model gives an accelerating universe and is consistent with current
observation\cite{Li:2004rb, Huang:2004ai}. For the recent works on holographic dark energy, see ref.
\cite{Zhang:2007sh,Sadjadi:2007ts, Saridakis:2007cy}. Recently, Gao et.al. \cite{Gao:2007ep} suggested that the dark
energy density may be inversely proportional to the Ricci scalar curvature, and they call this model the Ricci dark
energy model. They have shown that  the causality problem disappears in their model and the result is
phenomenologically viable. Since the Brans-Dicke theory is a natural extension of Einstein's general relativity and can
pass the experiments from the solar system \cite{Bertotti:2003rm}, it is worthwhile to investigate this model in the
Brans-Dicke theory. Therefore, in this letter, we study the evolution of the equation of state of the Ricci dark energy
and the transition from decelerated to accelerated expansion of the universe in the Brans-Dicke theory. For the
relevant works on holographic dark energy in Brans-Dicke theory, see ref. \cite{Gong:2004fq, Kim:2005rua, Kim:2005gk,
Setare:2006yj,Banerjee:2007zd,Nayak:2008pn,Xu:2008sn}.

This paper is organized as follows. In section 2 we briefly review the Brans-Dicke cosmology. In section 3 we study the
evolution of the equation of state of the Ricci dark energy and the transition from decelerated to accelerated
expansion of the universe. In the final section we will give some discussions and conclusions.
\section{Review on Brans-Dicke cosmology}
The Brans-Dicke theory of gravity is a natural extension of Einstein's general gravity. The action for the Brans-Dicke theory with a
perfect fluid \cite{Kim:2004is} in the Jordan frame is
\begin{equation}\label{BD action}
    S = \int d^4x\sqrt{g}\left[ \frac{1}{16\pi}\left( \Phi R - \omega \frac{\nabla_\mu \Phi \nabla^\mu\Phi}{\Phi} \right) + \mathcal {L}_M\right]
\end{equation}
where $\Phi$ is the Brans-Dicke scalar field representing the inverse of $Newton's$ constant which is allowed to vary
with space and time and $\omega$ is the generic dimensionless parameter of the Brans-Dicke theory. The Lagrangian
$\mathcal{L}_M$ represents the perfect fluid matter. In the Jordan frame, the matter minimally couples to the metric
and there is no interaction between the scalar field $\Phi$ and the matter field \cite{Gong:2004fq}. The equations of
motion for the metric $g_{\mu\nu}$ and the Brans-Dicke scalar field $\Phi$ are
\begin{equation}
    \begin{split}
       G_{\mu\nu} = R_{\mu\nu} - \ah g_{\mu\nu}R &= \frac{8\pi}{\Phi} T^M_{\mu\nu}+8\pi T^{BD}_{\mu\nu}\\
       \nabla_\mu \nabla^\mu \Phi &= \frac{8\pi}{2\omega + 3}T^{M \mu}_{\quad \mu}
    \end{split}
\end{equation}
Here the stress-energy-momentum tensor for the matter are defined as usual, namely $T^M_{\mu\nu} = (2/\sqrt{g})
\delta(\sqrt{g}\mathcal{L}_M) / \delta g^{\mu\nu}$. The explicit form of $T^M_{\mu\nu}$ is
\begin{equation}
    T^M_{\mu\nu} = (\rho_M + p_M)U_{\mu}U_{\nu} + p_M g_{\mu\nu}
\end{equation}
where $\rho_M$ and $p_M$ denotes the energy density and pressure of the matter respectively and $U_\mu$ is a four
velocity vector normalized as $U_\mu U^\mu = -1$ . However, the tress-energy-momentum tensor for the Brans-Dicke scalar
field $\Phi$ can not be defined in a similar way due to the non-minimal coupling term $\sqrt{g}\Phi R$ in the action
which obstructs the separation of the scalar $\Phi$ Lagrangian from the tensor $g_{\mu\nu}$ Lagrangian. The explicit
form of $T^{BD}_{\mu\nu}$ is
\begin{equation}
      T^{BD}_{\mu\nu} = \frac{1}{8\pi}\left[\frac{\omega}{\Phi^2}\left(\nabla_\mu \Phi \nabla_\nu\Phi - \ah g_{\mu\nu}\nabla_\alpha \Phi \nabla^\alpha \Phi \right)
      + \left(\nabla_\mu \nabla_\nu\Phi - g_{\mu\nu}\nabla_\alpha \nabla^\alpha \Phi\right) \right]
\end{equation}
Note that the Einstein' general relativity will be recovered in the $\omega \rightarrow \infty$ limit of the Brans-Dicke theory.
\\

Assuming our universe is homogeneous and isotropic on large scale, we work with the Friedmann- Robertson-Walker (FRW) spacetime
\begin{equation}\label{frwm}
    ds^2=-dt^2+a^2(t)\left[{dr^2\over1-kr^2}+r^2(d\theta^2+sin^2\theta d\phi^2)\right] \, ,
\end{equation}
where $a(t)$ is the scale factor of the universe. Here $k = -1, 0, 1$ represent that the universe is open, flat, closed respectively. For
simplicity we assuming $\Phi = \Phi(t)$, then in the FRW spaceime the field equations take the forms
\begin{equation}\label{FRW}
    H^2 + H\left( \frac{\dot\Phi}{\Phi} \right) - \frac{\omega}{6}\left( \frac{\dot\Phi}{\Phi} \right)^2 = \frac{8\pi}{3\Phi}\rho_M -
    \frac{k}{a^2}
\end{equation}
\begin{equation}\label{BD scalar}
    \ddot\Phi + 3H\dot\Phi = \frac{8\pi}{2\omega + 3}\left(\rho_M - 3p_M \right)
\end{equation}
\begin{equation}\label{Conservation}
    \dot\rho_M + 3H(\rho_M + p_M) = 0
\end{equation}
where $H = \dot a/ a$ represents the Hubble parameter and the overdot stands for the derivative with respect to the
cosmic time $t$.  The first equation (\ref{FRW}) corresponds to the Friedmann equation, the second equation (\ref{BD
scalar}) is  the equation of motion of the Brans-Dicke scalar field. The last is the conservation law $\nabla^\mu
T^M_{\mu\nu} = 0 $ for the matter fluid. We assume the Brans-Dicke scalar field can be regarded as a perfect fluid with
stress-energy-momentum tensor \cite{Kim:2004is}
\begin{equation*}
    \tilde T^{BD}_{\mu\nu} = (\rho_{BD} + p_{BD})U_{\mu\nu}U_{\mu\nu} + p_{BD}g_{\mu\nu}  \equiv T^{BD}_{\mu\nu}/G
\end{equation*}
where $G$ denotes the value of Newton's constant and its energy density and pressure are given by
\begin{equation*}
    \begin{split}
        \rho_{BD} &= \frac{1}{16\pi G} \left[ \omega\left( \frac{\dot\Phi}{\Phi}\right)^2 - 6H\left( \frac{\dot\Phi}{\Phi}\right) \right] \\
        p_{BD} &= \frac{1}{16\pi G} \left[ \omega\left( \frac{\dot\Phi}{\Phi}\right)^2 + 4H\left( \frac{\dot\Phi}{\Phi}\right) + 2 \frac{\ddot\Phi}{\Phi}\right]
    \end{split}
\end{equation*}
Finally, the Bianchi identity $\nabla^\mu G_{\mu\nu} = 0 $ , which plays the consistency relation \cite{Kim:2005rua,
Kim:2005gk, Kim:2004is}, leads to an energy transfer between the Brans-Dicke field and other matter.
\begin{equation}\label{consistent}
    \dot \rho_{BD} + 3H(\rho_{BD} + p_{BD}) = \frac{\dot\Phi}{\Phi^2 G} \rho_M  \, .
\end{equation}
Any physical solutions should satisfy these equations (\ref{FRW})-(\ref{consistent}) simultaneously.

\section{Ricci dark energy in BD theory}

Recently, Gao et.al \cite{Gao:2007ep} have proposed a holographic dark energy model in which the future event horizon
is replaced by the inverse of the Ricci scalar curvature. This model does not only avoid the causality problem and is
phenomenologically viable, but also solve the coincidence problem of dark energy. The Ricci curvature of FRW universe
is given by
\begin{equation}\label{Ricci}
    R = -6(\dot H + 2H^2 + \frac{k}{a^2}) \, .
\end{equation}
They introduced a holographic dark energy proportional to the Ricci scalar
\begin{equation}\label{Ricci DE}
    \rho_\Lambda = \frac{3\alpha}{8\pi G} \left(\dot H + 2H^2 + \frac{k}{a^2}\right) \propto R
\end{equation}
where the dimensionless coefficient $\alpha$ will be determined by observations and they call this model the Ricci dark
energy model. Solving the Friedmann equation they find the result
\begin{equation}
   \Omega_{\Lambda} \equiv \frac{8\pi G }{3H^2_0}\rho_\Lambda  = \frac{\alpha}{2-\alpha}\Omega_{m0}e^{-3x} + f_0e^{-(4-\frac{2}{\alpha})x}
\end{equation}
where $\Omega_{m0} \equiv 8\pi G\rho_{m0}/3H^2_0$, $x = \ln{a}$ and $f_0$ is an integration constant. Taking the
observation values of parameters they find the $\alpha \simeq 0.46 $ and $f_0 \simeq 0.65$. The evolution of the
equation of state $w_{\Lambda} \equiv p_\Lambda / \rho_\Lambda $ of dark energy is the following. At high redshifts the
value of $w_{\Lambda} $ is closed to zero, namely the dark energy behaves like the cold dark matter, and n,owadays
$w_{\Lambda} $ approaches $-1$ as required and in the future the dark energy will be phantom. Further more this model
can avoid the age problem and the causality problem.

It is worthwhile to investigate the Ricci dark energy in the framework of the Brans-Dicke theory. Note that in
Brans-Dicke theory, the scalar field $\Phi$ plays the role of Newton's constant $( \Phi\sim 1/G)$, so it is natural to
modify Ricci dark energy (\ref{Ricci DE}) in the Brans-Dicke theory as the following
\begin{equation}\label{m Ricci DE}
    \rho_\Lambda = \frac{3\alpha \Phi}{8\pi} \left(\dot H + 2H^2 + \frac{k}{a^2}\right) .
\end{equation}
The energy density of the perfect fluid matter $\rho_M$ in (\ref{FRW}) contains the cold dark matter $\rho_m$,
radiation $\rho_\gamma$ and the Ricci dark energy $\rho_\Lambda$. Every component itself satisfies the conservation law
(\ref{Conservation}). Let us assume the Brans-Dicke field vary with time as a power law of the scale factor
\cite{Nayak:2008pn}, and let $x = \ln a$
\begin{equation}\label{power}
    \Phi(t) = \Phi_0 e^{nx} = \frac{1}{G}e^{nx}
\end{equation}
Here we set the present scale factor $a_0 = 1$, so $\Phi_0 = 1/G$. One can see that in order to be consistent with
observations on Newton constant, the value of $n$ should be small. Then eq.(\ref{FRW}) becomes
\begin{equation}\label{FRW2}
H^2(1 + n - \frac{\omega}{6} n^2) = \frac{8\pi G}{3}\rho_M e^{-nx} - ke^{-2x}
\end{equation}
and the Ricci dark energy can be rewritten as
\begin{equation}\label{Ricci DE re}
     \rho_\Lambda = \frac{3\alpha}{8\pi G} \left( \ah (H^2)' + 2H^2 + k e^{-2x}\right)e^{nx} \, ,
\end{equation}
where prime denotes the derivative with respect to $x$. The energy density of cold dark matter and radiation are solved  by conservation
laws
\begin{equation}\label{CM and rad}
    \rho_m = \rho_{m0} e^{-3x}  \, ,\quad \rho_\gamma = \rho_{\gamma0} e^{-4x}
\end{equation}
where $\rho_{m0}$ and $\rho_{\gamma0}$ are the present energy density of matter and radiation respectively. Putting eq.(\ref{Ricci DE re}),
(\ref{CM and rad}) in eq.(\ref{FRW2}), we get
\begin{equation}\label{FRW3}
\begin{split}
H^2(1 + n - \frac{\omega}{6} n^2) &= \alpha\left(  \ah (H^2)' + 2H^2  \right) + (\alpha - 1)ke^{-2x} \\
                                   &+ H_0^2\left( \Omega_{m0} e^{-(3+n)x} +  \Omega_{\gamma0} e^{-(4+n)x}\right) \\
\end{split}
\end{equation}
where $H_0$ is the present Hubble parameter. Solving the eq.(\ref{FRW3}) we get
\begin{equation}\label{FRW4}
\begin{split}
    H^2 &= \frac{\alpha-1}{A-\alpha}ke^{-2x}
    + \frac{2H_0^2}{(-\alpha+2A+\alpha n)}\Omega_{m0}e^{-(3+n)x}\\
    &+\frac{2H_0^2}{(2A+\alpha n)}\Omega_{\gamma 0}e^{-(4+n)x}+g_0H^2_0e^{-\frac{2(2\alpha-A)x}{\alpha}}
\end{split}
\end{equation}
where $A\equiv 1 + n - \frac{\omega}{6}n^2$ and $g_0$ is an integration constant determined later. Then the energy density of the Ricci
dark energy is
\begin{equation}\label{Ricci DE sol}
\begin{split}
    \Omega_\Lambda &\equiv\frac{8\pi G}{3H^2_0}\rho_\Lambda =\frac{\alpha(1-n)}{2A+\alpha(n-1)}\Omega_{m0}e^{-3x}
    - \frac{\alpha n}{2A + \alpha n}\Omega_{\gamma0}e^{-4x}\\
    &\\
    & +\frac{(A-1)\alpha}{A-\alpha}\Omega_{k0}e^{(n-2)x}
    + g_0 A e^{(n-4 + \frac{2A}{\alpha})x}
\end{split}
\end{equation}
where $\Omega_{k0}\equiv k/H^2_0$. Using the conservation law,
\begin{equation}\label{pressure cl}
  \rho_\Lambda' + 3(\rho_\Lambda + p_\Lambda ) = 0
\end{equation}
we get the pressure of the dark energy
\begin{equation}\label{pressure}
\begin{split}
   \frac{8\pi G}{H^2_0} p_\Lambda &= -\frac{\alpha n }{2A + \alpha n}\Omega_{\gamma0}e^{-4x}
    -  \frac{\alpha(n+1)(A-1)}{A-\alpha}\Omega_{k0}e^{(n-2)x}\\
    &\\
    &- \left(n-1+\frac{2A}{\alpha}\right)g_0 A e^{(n-4 + \frac{2A}{\alpha})x}\\
\end{split}
\end{equation}

Notice that the equation of motion of the Brans-Dicke scalar field (\ref{BD scalar}) can be also rewritten in terms of $x$
\begin{equation}\label{BD scalar 2}
   H^2 (n^2 + 3n) + \ah n(H^2)' = \frac{ 3H^2_0e^{-nx}}{(2w + 3)}\left(\Omega_{m0}e^{-3x} + \frac{8\pi G}{3H^2_0}\rho_\Lambda - \frac{8\pi G}{H^2_0}p_\Lambda
   \right)
\end{equation}
Putting eq.(\ref{FRW4}), (\ref{Ricci DE sol}) and (\ref{pressure}) in eq.(\ref{BD scalar 2}), we obtain the equation
for $n$
\begin{equation}\label{eqn}
\begin{split}
    &(n+2)\left(n\frac{\alpha -1}{A - \alpha}- \frac{3\alpha}{2\omega +3}\frac{A-1}{A-\alpha}\right)\Omega_{k0}e^{-2x}\\
    &\\
    &+\left(\ah n(n+3) -\frac{3A}{2\omega +3} \right)\frac{2\Omega_{m0}e^{-(3+n)x}}{2A+\alpha(n-1)}
    +n \left(\frac{n}{2} +1\right)\frac{2\Omega_{\gamma0}e^{-(4+n)x}}{2A+\alpha n}\\
    &\\
    &+\left( n(n+1+\frac{A}{\alpha}) - \frac{3}{2\omega +3}(n+\frac{2A}{\alpha})\right)g_0e^{-(4-\frac{2A}{\alpha})x} = 0\\
\end{split}
\end{equation}
Local astronomical experiments set a very high lower bound on $\omega$ \cite{Will:1993ns} The Cassini experiment
\cite{Bertotti:2003rm} implies that $\omega > 10^4$. Likewise, a slow fractional variation of $\Phi$ will lead to a
small fraction variation of $G$, consistent with observations. Therefore, the interesting case is that when $|n|$ is
small whereas $\omega$ is large so that the product $n^2\omega$ results of order unity \cite{Banerjee:2007zd}, namely
in the limit of $\omega \rightarrow \infty$ and $n^2\omega \simeq O(1)$. So $A \simeq 1 - \frac{1}{6}n^2\omega$ and
eq.(\ref{eqn}) becomes
\begin{equation}\label{eqn2}
    \begin{split}
    &\frac{2(\alpha -1)\Omega_{k0}}{1 - \frac{1}{6}n^2\omega - \alpha}ne^{-2x}
    +\frac{3\Omega_{m0}}{(2-\frac{1}{3}n^2\omega-\alpha)}ne^{-3x}\\
    &+\frac{2\Omega_{\gamma0}}{(2-\frac{1}{3}n^2\omega)}ne^{-4x}
    +g_0\left(1+\frac{1}{\alpha} - \frac{n^2\omega}{6\alpha} \right)ne^{-(4-\frac{2}{\alpha}+\frac{n^2\omega}{3\alpha})x} = 0\\
\end{split}
\end{equation}
We see that as long as $n$ is small enough, i.e. smaller than the exponential factor, the above equation is always
approximately satisfied. But, if the exponential factor becomes very large, the above equation will be no longer
satisfied, then our results will be invalid. In the following we assume that this equation has a exact solution at $x =
x_0$,
\begin{equation}\label{eqn3}
\begin{split}
    &\frac{2(\alpha -1)\Omega_{k0}e^{-2x_0}}{1 - \frac{1}{6}n^2\omega - \alpha}
    +\frac{3\Omega_{m0}e^{-3x_0}}{(2-\frac{1}{3}n^2\omega-\alpha)}\\
    &\\
    &+\frac{2\Omega_{\gamma0}e^{-4x_0}}{(2-\frac{1}{3}n^2\omega)}
    +g_0\left(1+\frac{1}{\alpha} - \frac{n^2\omega}{6\alpha} \right)e^{-(4-\frac{2}{\alpha}+\frac{n^2\omega}{3\alpha})x_0} = 0\\
\end{split}
\end{equation}
We will use this equation to constraint our parameters defined later.

One can check that the consistent condition eq.(\ref{consistent}) written in terms of $x$ is
\begin{equation}\label{consistent x}
\begin{split}
    &6H^2(n^2\omega - n + n^2) + (H^2)'(n^2\omega - 3n)\\
    &\\
    = &\, 6H^2_0 n e^{-nx}(\Omega_\Lambda + \Omega_{m0}e^{-3x} + \Omega_{\gamma0}e^{-4x})\\
\end{split}
\end{equation}
Putting eq.(\ref{FRW4}) and (\ref{Ricci DE sol})  in eq.(\ref{consistent x}), we obtain the equation for $n$
\begin{equation}\label{consistent x2}
\begin{split}
    &\left[(-n^2\omega + 5n + n^2)\frac{\alpha -1}{A-\alpha}-6n\alpha\frac{A-1}{A-\alpha}\right]\Omega_{k0}e^{-2x}\\
    &\\
    &+\left[ \frac{2(-2n^2\omega + 8n + 4n^2 - n^3\omega)}{-\alpha + 2A + \alpha n} -\frac{12nA}{2A+\alpha(n-1)}\right]\Omega_{m0}e^{-(3+n)x}\\
    &\\
    &+\left[ \frac{2(-3n^2\omega + 11n + 4n^2 - n^3\omega)}{2A+\alpha n}-\frac{12nA}{2A+\alpha n}\right]\Omega_{\gamma0}e^{-(4+n)x}\\
    &\\
    &+\left[ (\frac{2A}{\alpha}-3)n^2\omega + n^2 + (11 - \frac{6A}{\alpha} - 6A)n\right]g_0e^{(-4 + \frac{2A}{\alpha})x} = 0\\
\end{split}
\end{equation}
Taking the same limit as before, we get
\begin{equation}\label{consistent limit}
\begin{split}
    &\frac{n^2\omega(1-\alpha)}{1-\frac{n^2\omega}{6}-\alpha}\Omega_{k0}e^{-2x}
    -\frac{4n^2\omega}{2-\frac{n^2\omega}{3}-\alpha } \Omega_{m0}e^{-3x}\\
    &-\frac{6n^2\omega}{2-\frac{n^2\omega}{3}}\Omega_{\gamma0}e^{-4x}
    +(\frac{2}{\alpha}-\frac{n^2\omega}{3\alpha}-3)n^2\omega g_0e^{(-4 + \frac{2}{\alpha}-\frac{n^2\omega}{3\alpha})x} = 0\\
\end{split}
\end{equation}
Since $n^2\omega$ is of order unity, the above equation can not be always satisfied, so we assume this equation is valid near $x=0$, which
means it is satisfied nowadays, i.e. when $a=a_0 = 1$
\begin{equation}\label{consistent now}
\begin{split}
    &\frac{(1-\alpha)\Omega_{k0}}{1-\frac{n^2\omega}{6}-\alpha}
    -\frac{4 \Omega_{m0}}{2-\frac{n^2\omega}{3}-\alpha }
    -\frac{6\Omega_{\gamma0}}{2-\frac{n^2\omega}{3}}
    +\left(\frac{2}{\alpha}-\frac{n^2\omega}{3\alpha}-3\right) g_0 = 0\\
\end{split}
\end{equation}
In fact, one shall see that the above equation is really satisfied. Notice that this equation can not give constraint on the parameters we
defined later, it is just used for consistent check.

Let's come back to the energy density of the dark energy in eq.(\ref{Ricci DE sol}). Taking the same limit and keeping the first order of
$n$,  we obtain
\begin{equation}\label{Ricci DE sol limit}
\begin{split}
    \Omega_\Lambda &=\frac{\alpha}{2-\frac{1}{3}n^2\omega-\alpha}\Omega_{m0}e^{-3x} - \frac{\alpha n }{2-\frac{1}{3}n^2\omega-\alpha}\Omega_{m0}e^{-3x}
    - \frac{\alpha n}{2-\frac{1}{3}n^2\omega}\Omega_{\gamma0}e^{-4x} \\
    & +  \frac{n^2\omega\alpha}{n^2\omega + 6\alpha - 6}\Omega_{k0}e^{-2x}
    + g_0 (1-\frac{1}{6}n^2\omega)e^{-(4-\frac{2}{\alpha}+\frac{n^2\omega}{3\alpha})x}
\end{split}
\end{equation}
Here we can see that when $n$ is exactly zero and $\omega$ is finite, the energy density comes back to the result in \cite{Gao:2007ep}.
Compare with their result, the contribution of radiation on the right hand side of the above equation is only the term with the first order
of $n$, which means in the small $n$ limit, the radiation part is not important. In another word, radiation contribution to the energy
density is the high order effect compare to the matter part. The matter terms on the right hand side of eq.(\ref{Ricci DE sol limit})
consist of two parts, one for zeroth order of $n$ and the other the first order.  For simplicity we just keep the terms with zeroth order
of $n$, namely
\begin{equation}\label{Ricci DE sol limit zero}
\begin{split}
    \Omega_\Lambda =C_0\Omega_{m0}e^{-3x}
    + D_0\Omega_{k0}e^{-2x}
    + F_0e^{\left(-3 + C_0^{-1}\right)x}
\end{split}
\end{equation}
where we have define  $C_0 \equiv \frac{\alpha}{2-\frac{1}{3}n^2\omega-\alpha}$, $D_0 \equiv \frac{n^2\omega\alpha}{n^2\omega + 6\alpha -
6} $ and $F_0 \equiv g_0 (1-\frac{1}{6}n^2\omega)$.

Now we can rewritten the equation of motion (\ref{eqn3}) with these three parameters as follows
\begin{equation}\label{eqm3}
    \begin{split}
    &\frac{4C_0(1-D_0)\Omega_{k0}e^{-2x_0} + 3C_0(1+C_0)\Omega_{m0}e^{-3x_0} + 2C_0\Omega_{\gamma0}e^{-4x_0} }{2C_0 - (C_0 -1)D_0}\\
    &\\
    &+\frac{F_0(1+3C_0)e^{(-3+C_0^{-1})x_0}}{2C_0 - (C_0 -1)D_0}= 0\\
\end{split}
\end{equation}
and the consistent equation (\ref{consistent now}) can be also rewritten in terms of $C_0$, $D_0$ and $F_0$.
\begin{equation}\label{consistent now2}
\begin{split}
    \frac{2C_0\Big[(1-D_0)\Omega_{k0} - 2(1+C_0) \Omega_{m0} - 3\Omega_{\gamma0}\Big] + 2F_0(1-2C_0)}{2C_0-(C_0-1)D_0} = 0
\end{split}
\end{equation}
where we have used
\begin{equation}\label{trans}
    \alpha = \frac{2C_0 - (C_0-1)D_0}{1+C_0}, \quad n^2\omega = 3\left(1-C_0^{-1}\right)D_0
\end{equation}
and
\begin{equation}\label{transg}
    g_0 =\frac{2F_0}{2-(1-C_0^{-1})D_0}
\end{equation}
The pressure of the dark energy in eq.(\ref{pressure}) can be also rewritten in terms of these three constants and in the same limit as
before
\begin{equation}\label{pressure3}
\begin{split}
   \frac{8\pi G}{H^2_0} p_\Lambda &=
    -  \frac{n^2\omega\alpha}{n^2\omega+6\alpha-6}\Omega_{k0}e^{-2x}
    - \left(\frac{2}{\alpha}-\frac{n^2\omega}{3\alpha}-1\right)g_0 \left(1-\frac{n^2\omega}{6}\right) e^{\left(-3 + C_0^{-1}\right)x}\\
    &\\
    &= -D_0\Omega_{k0}e^{-2x} - C_0^{-1}F_0e^{\left(-3 + C_0^{-1}\right)x}
\end{split}
\end{equation}
Defining the equation of state of the dark energy $w_\Lambda = p_\Lambda/\rho_\Lambda$, we get
\begin{equation}\label{equation of state}
    w_\Lambda(x) = -\frac{1}{3\Omega_\Lambda}\left[D_0\Omega_{k0}e^{-2x} + C_0^{-1}F_0e^{\left(-3 + C_0^{-1}\right)x}\right]
\end{equation}

There are three parameters $C_0$, $D_0$ and $F_0$ to be determined by eq.(\ref{Ricci DE sol limit zero}),(\ref{eqm3}) and (\ref{equation of
state})in terms of the present values of cosmological parameters as follows
\begin{equation}\label{determinate}
\begin{split}
    \Omega_{\Lambda0} &= C_0\Omega_{m0} + D_0\Omega_{k0} + F_0 \\
    &\\
    w_{\Lambda0} &=-\frac{1}{3\Omega_{\Lambda0}}\left[D_0\Omega_{k0} + C_0^{-1}F_0\right]\\
    &\\
    4(1-D_0)\Omega_{k0}e^{-2x_0} &+ 3(1+C_0)\Omega_{m0}e^{-3x_0}  + 2\Omega_{\gamma0}e^{-4x_0}  \\
    &+ (C_0^{-1}+3)F_0e^{(-3+C_0^{-1})x_0} =0\\
\end{split}
\end{equation}
and the consistent equation
\begin{equation}\label{consistent check}
        \Omega_{k0} - 2\Omega_{m0} - 3\Omega_{\gamma0}-3w_{\Lambda0}\Omega_{\Lambda0}-2\Omega_{\Lambda0} = 0
\end{equation}
Note that when $\Omega_{k0}\sim 0$, $\Omega_{\gamma0}\sim 0$ and $\Omega_{\Lambda0}\sim3\Omega_{mo}$ , this consistent equation requires
$w_\Lambda \sim -0.9$ which is observational accepted, in other words, this consistent equation is really satisfied at present day. Then we
get
\begin{equation}\label{cal para}
\begin{split}
    C_0 &= \frac{\Omega_{\Lambda0} - D_0\Omega_{k0}}{\Omega_{m0}-3w_{\Lambda0}\Omega_{\Lambda0}-D_0\Omega_{k0}}\\
    &\\
    F_0 &= -C_0(3w_{\Lambda0}\Omega_{\Lambda0} + D_0\Omega_{k0})\\
\end{split}
\end{equation}
Here we can use eq.(\ref{cal para}) to eliminate $C_0$ and $F_0$ in the last equation of (\ref{determinate}). Then we get the last equation
as equation of $D_0$. In fact this equation contains another parameters $x_0$, and both $D_0$ and $x_0$ determines the zero point of the
function on the left hand side of this equation. So we have the freedom to choice $D_0$ or $D_0\Omega_{k0}$, and solve the value of $x_0$
to satisfy the equation. Generally, the value of $x_0$ cannot be zero, which means the equation is not exactly satisfied presently, but one
should not forget that actually  we have a very small factor $n$ multiplied on the left side of the last equation of (\ref{determinate}),
see eq.(\ref{eqn2}). So that even though this equation is not exactly satisfied at $x = 0$, i.e., at $a=a_0 =1$, it can be approximately
satisfied as long as $n$ is small enough. When $\Omega_{k0} = 0$, the last equation in (\ref{determinate}) does not contain the $D_0$
parameters any more, and the value of $x_0$ will change, but again small enough $n$ will make the equation approximately satisfied at
present. Define a function of $C_0$, $D_0$ and $F_0$ as
\begin{equation}\label{boundfun}
    N^{-1}(D_0\Omega_{k0}) = 4(1-D_0)\Omega_{k0}+ 3(1+C_0)\Omega_{m0}  + 2\Omega_{\gamma0}+ (C_0^{-1}+3)F_0
\end{equation}
Then $n$ should be much less than the value of $N$, namely $n << N$, to make the equation of motion (\ref{eqn2}) approximately satisfied at
$x = 0$. Note that we can use eq.(\ref{cal para}) to eliminate $C_0$ and $F_0$ in eq.(\ref{boundfun}).

As an example we will consider the following values of cosmological parameters \cite{Spergel:2006hy} :
$\Omega_{\Lambda0} = 0.72$, $\Omega_{m0}=0.27$, $\Omega_{\gamma0} = 10^{-4}$, $\Omega_{k0} = 10^{-3}$, $w_{\Lambda0} =
-1$. We draw a  3D graphic to show the evolution of the equation of state $w_{\Lambda0} $ with  parameter
$D_0\Omega_{k0}$ varying in Fig.1.
\\

\bigskip{
    \vbox{
            {
                \nobreak
                \centerline
                {
                    \includegraphics[scale=0.8]{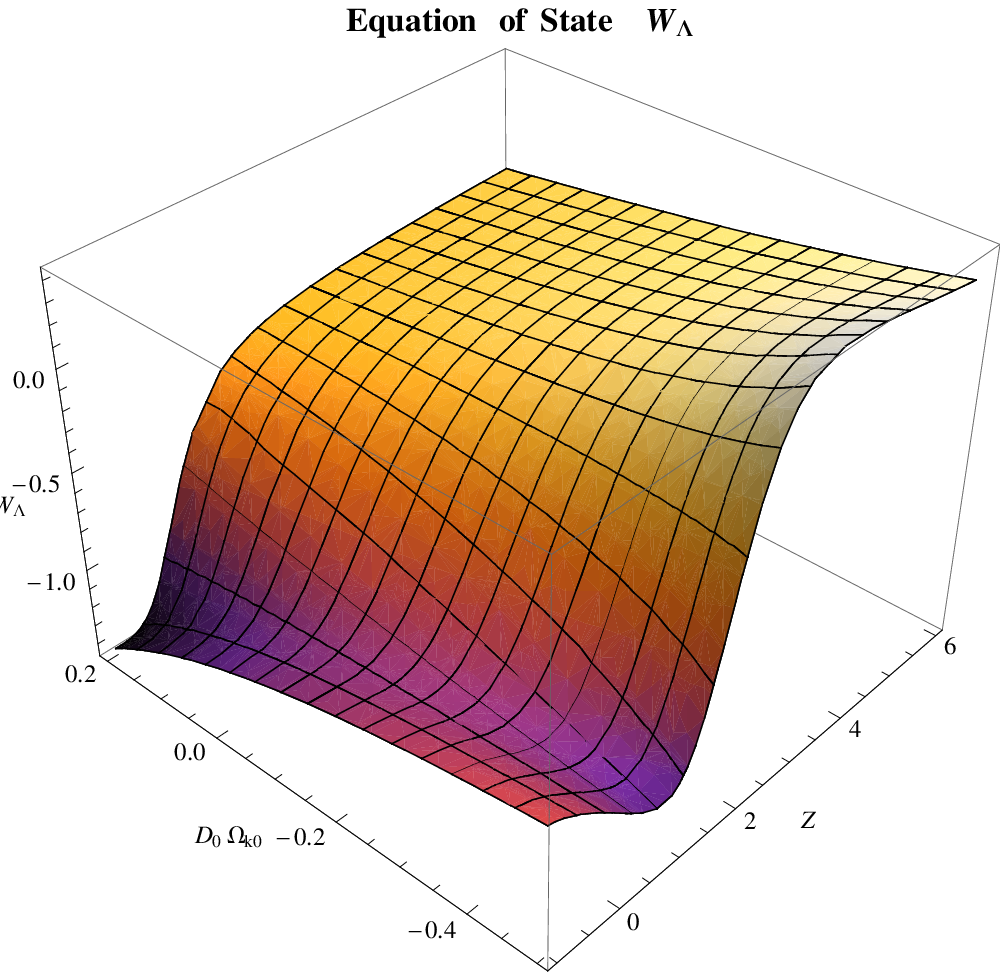}
                }
                \nobreak
                \bigskip
                {\raggedright\it \vbox
                    {
                        {\bf Figure 1.}
                        {\it The evolution of the equation of state $w_{\Lambda0} $ with  parameter $D_0\Omega_{k0}$ varying. The redshift $z$ is
                             defined as $a = (1+z)^{-1}$.
                        }
                    }
                }

            }
        }
\bigskip}
\noindent Here $z = 0$ corresponds to the present day. We also choice some typical values of $D_0\Omega_{k0}$ to plot the evolution of
$w_\Lambda$ in Fig.2
\\

\bigskip{
    \vbox{
        {
            \nobreak
            \centerline
            {
                \includegraphics[scale=0.8]{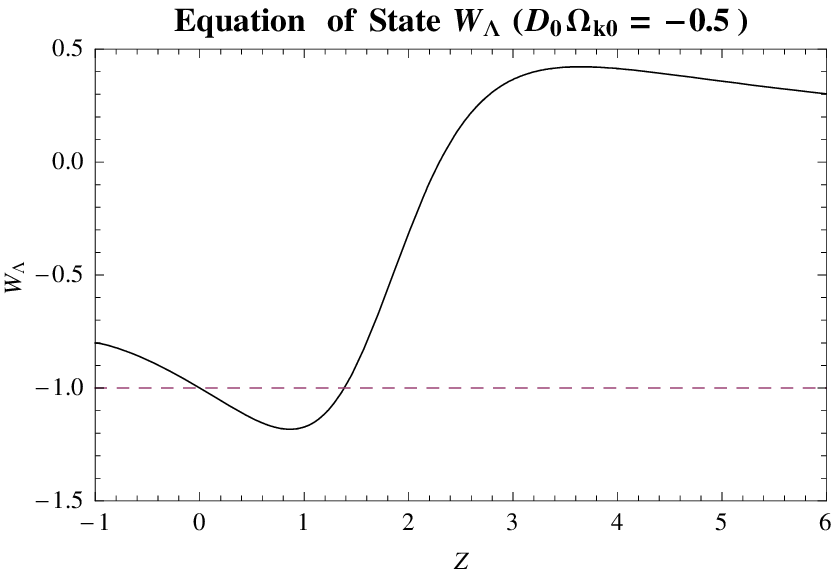}
                \includegraphics[scale=0.8]{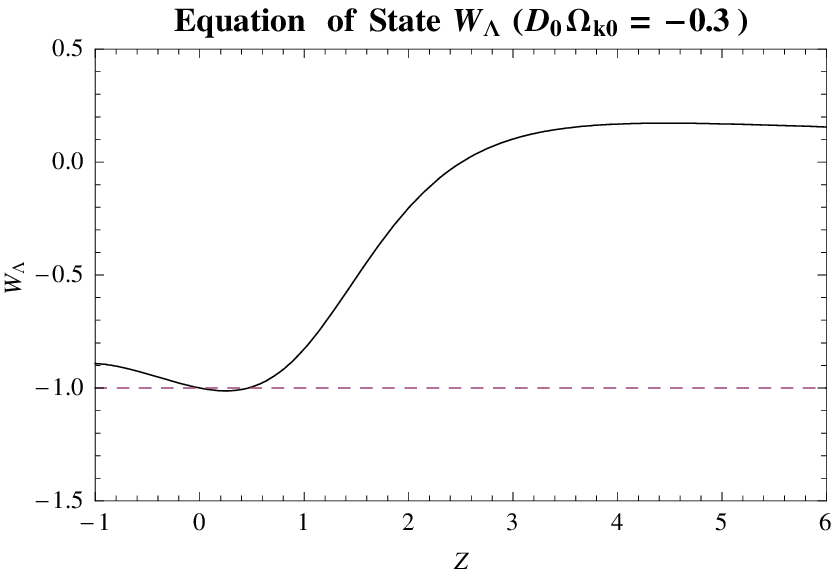}
            }
            \nobreak
        }

    }
\bigskip}

\bigskip{
    \vbox{
        {
            \nobreak
            \centerline
            {
                \includegraphics[scale=0.8]{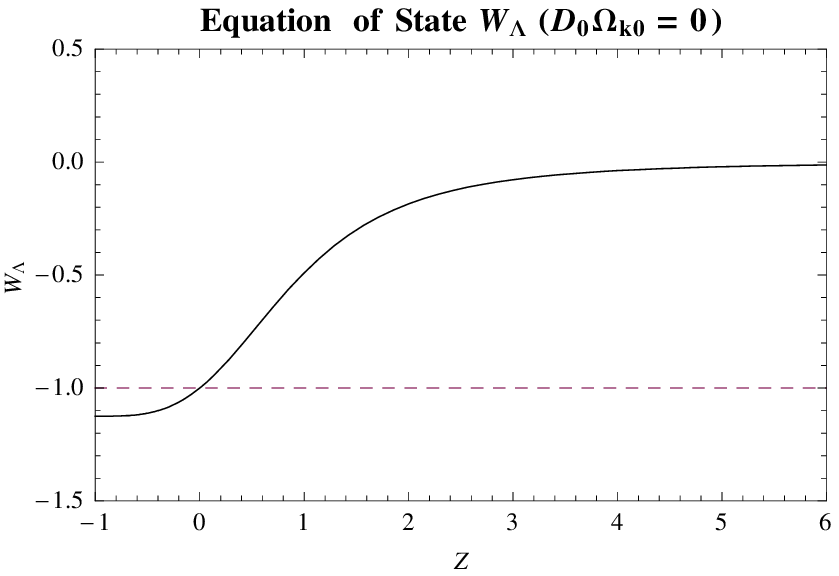}
                \includegraphics[scale=0.8]{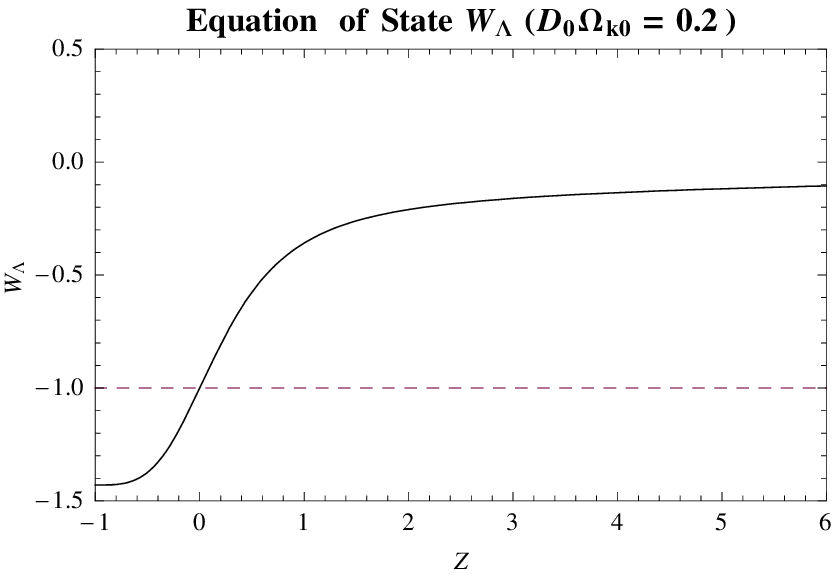}
            }
            \nobreak
            \bigskip
            {
                \raggedright \it \vbox
                {
                    {\bf Figure 2.}
                    {\it The evolution of the equation of state $w_{\Lambda0} $ with  different values of parameter $D_0\Omega_{k0}$.
                         The dashed line corresponds to $w_\Lambda = -1$.
                    }
                }
            }
        }

    }
\bigskip}
\noindent One can see that when $D_0\Omega_{k0} = 0$, the result is the same as that in ref.\cite{Gao:2007ep} as follows. The equation of
state is nearly zero at high redshifts, so the dark energy behaves like matter.  $w_\Lambda$ approaches $-1$ at $z\sim0$, and in the future
$w_\Lambda$ will be less than $-1$. When $D_0\Omega_{k0} \sim 0.2$ the whole curve is almost the same as that when $D_0\Omega_{k0} = 0$.
The interesting things is that when $D_0\Omega_{k0} \sim -0.3$, $w_\Lambda$ will never less than $-1$, and when $D_0\Omega_{k0} \sim -0.5$,
$w_\Lambda$ cross $-1$ twice.

As we discussed before the power $n$ in eq.(\ref{power}) should be much less than the function $N$ defined in (\ref{boundfun}) to make the
equation of motion satisfied at present day. We plot the values of $N$ with respect to $D_0\Omega_{k0}$ in Fig.3. One can see from the
figure that as long as $n << 0.1$ , which is often the case, $n$ is much less than $N$.
\\

\bigskip{
    \vbox{
        {
            \nobreak
            \centerline
            {
                \includegraphics[scale=1]{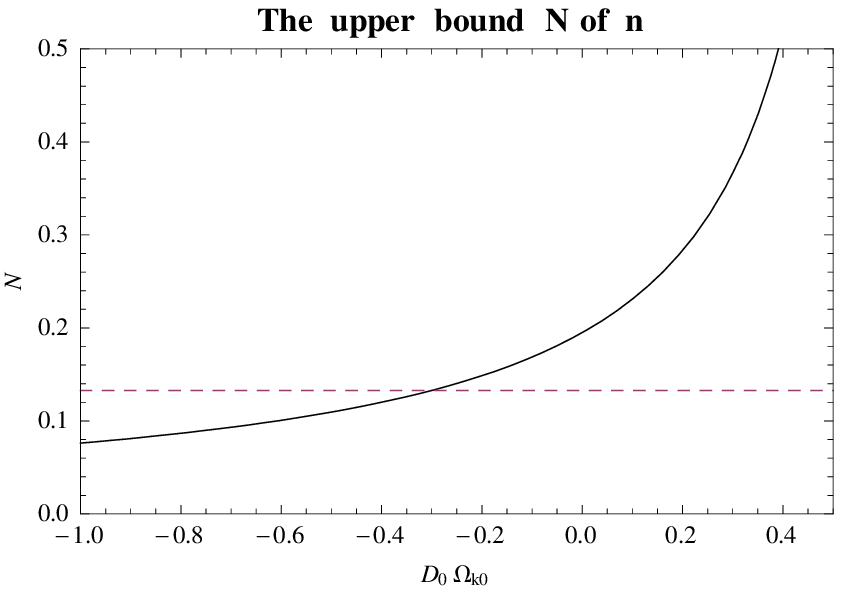}
            }
            \nobreak
            \bigskip
            {
                \raggedright \it \vbox
                {
                    {\bf Figure 3.}
                    {\it The upper bound $N(D_0\Omega_{k0})$ of $n$  with  different values of parameter $D_0\Omega_{k0}$.
                         The dashed line corresponds to $N(D_0\Omega_{k0}=-0.3)$.
                    }
                }
            }
        }

    }
\bigskip}
In Fig.4 we plot the evolution of the deceleration parameter defined as
\begin{equation}\label{dec par}
    q \equiv -\frac{\ddot a}{aH^2}
\end{equation}
which can be rewritten by
\begin{equation}\label{dec par 2}
    q = -1 - \frac{\dot H}{H^2} = -1 - \frac{H'}{H}
\end{equation}

\bigskip{
    \vbox{
        {
            \nobreak
            \centerline
            {
                \includegraphics[scale=1]{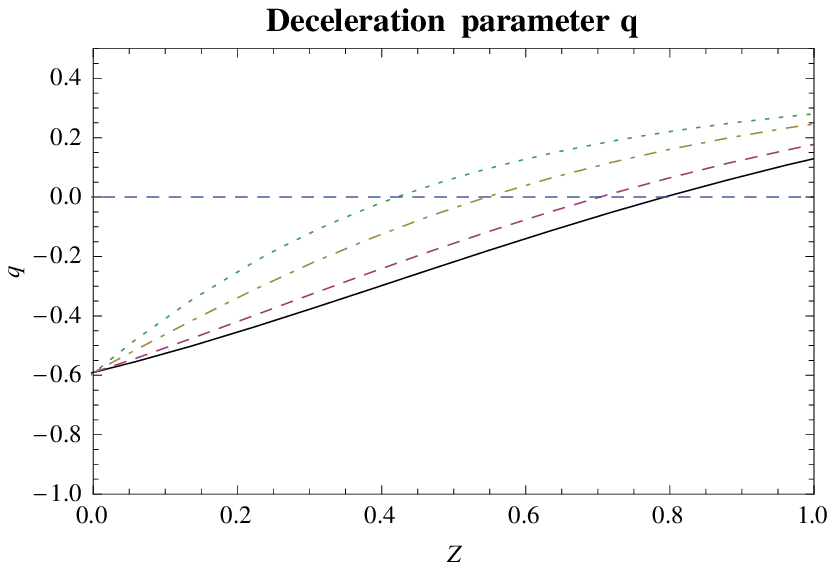}
            }
            \nobreak
            \bigskip
            {
                \raggedright \it \vbox
                {
                    {\bf Figure 4.}
                    {\it The deceleration parameter $q$ with  different values of parameter $D_0\Omega_{k0}$.
                         The horizontally dashed line corresponds to $q = 0$. The solid, dashed, dotdashed and dotted curves represent $D_0\Omega_{k0} = -0.5$, $-0.3$, $0$,
                         and $0.2$ respectively.
                    }
                }
            }
        }

    }
\bigskip}

\noindent  From Fig.4,  one can see that our universe speeds up as $z\simeq 0.4\sim 0.8$ for different values of $D_0\Omega_{k0}$. This is
consistent with the joint analysis of SNe $+$ CMB data $z_T = 0.52\sim 0.73$ \cite{Alam:2004jy,Guo:2006ce}.

\section{Discussion and Conclusion}
We have used different values of $D_0\Omega_{ko}$ to study the evolution of equation of state of the Ricci dark energy
and the transition from decelerated to accelerated expansion of the universe in the framework of Brans-Dicke theory.
The parameter $D_0\Omega_{k0}$  should be determined by further observations. When $D_0$ or $\Omega_{k0}$ vanished, we
come back to the result in ref.\cite{Gao:2007ep}, which means the role of the Brans-Dicke scalar field played can be
ignored when spatial curvature vanished. Once $D_0\Omega_{k0}$ does not equal to zero, the evolution of the equation of
state of the dark energy will be changed. Of course, the large absolute value of $D_0\Omega_{k0}$ are ruled out by
current observation. We also find that the contribution of radiation is from the next order of the small quantity $n$
which comes from the assumption of the power law of the Brans-Dicke scalar fields, so that it can be ignored
approximately.

\section*{ACKNOWLEDGEMENTS}
The author would like to thank Miao Li for a careful reading of the manuscript and valuable suggestions. We are
grateful to Yushu Song, Tower Wang, Yi Wang for useful discussions.

\end{document}